\documentclass[fleqn,usenatbib]{mnras}

\usepackage{newtxtext,newtxmath}
\usepackage[T1]{fontenc}

\DeclareRobustCommand{\VAN}[3]{#2}
\let\VANthebibliography\thebibliography
\def\thebibliography{\DeclareRobustCommand{\VAN}[3]{##3}\VANthebibliography}

\usepackage{graphicx}	% Including figure files
\usepackage{amsmath}	% Advanced maths commands
\title[THz DES of space relevant ices]{TeraHertz Desorption Emission Spectroscopy (THz DES) of space relevant ices}

\author[O. Auriacombe et al.]{
Olivier Auriacombe,$^{1,2}$\thanks{E-mail: o.auriacombe@gmail.com}
S. Rea,$^{2}$
S. Ioppolo,$^{3}$
M. Oldfield,$^{2}$
S. Parkes,$^{4}$
B. Ellison$^{2}$
and H. Fraser$^{1}$
\\
$^{1}$School of Physical Sciences, The Open University, Walton Hall, Milton Keynes, MK7 6AA, UK\\
$^{2}$Millimetre Wave Technology Group, STFC Rutherford Appleton Laboratory, Didcot, OX11 0QX, UK\\
$^{3}$School of Electronic Engineering and Computer Science, Queen Mary University of London, London, E14NS, UK\\
$^{4}$STAR-Dundee Ltd., Dundee, DD1 4EE, UK
}

\date{Accepted XXX. Received YYY; in original form ZZZ}

\pubyear{2022}

\begin{document}
\label{firstpage}
\pagerange{\pageref{firstpage}--\pageref{lastpage}}
\maketitle

\begin{abstract}
We present an experimental instrument that performs laboratory-based gas-phase Terahertz Desorption Emission Spectroscopy (THz-DES) experiments in support of astrochemistry. The measurement system combines a terahertz heterodyne radiometer that uses room temperature semiconductor mixer diode technology previously developed for the purposes of Earth observation, with a high-vacuum desorption gas cell and high-speed digital sampling circuitry to enable high spectral and temporal resolution spectroscopy of molecular species with thermal discrimination. During use, molecules are condensed onto a liquid nitrogen cooled metal finger to emulate ice structures that may be present in space. Following deposition, thermal desorption is controlled and initiated by means of a heater and monitored via a temperature sensor. The ‘rest frequency’ spectral signatures of molecules released into the vacuum cell environment are detected by the heterodyne radiometer in real-time and characterised with high spectral resolution. To demonstrate the viability of the instrument, we have studied Nitrous Oxide ($N_2O$). This molecule strongly emits within the terahertz (sub-millimetre wavelength) range and provide a suitable test gas and we compare the results obtained with more traditional techniques such as quadrupole mass spectrometry. The results obtained allow us to fully characterize the measurement method and we discuss its potential use as a laboratory tool in support of astrochemical observations of molecular species in the interstellar medium and the Solar System.
\end{abstract}

\begin{keywords}
Astrochemistry, Molecular processes, Radiative transfer, Instrumentation: detectors, Methods: laboratory: molecular, Techniques: spectroscopic
\end{keywords}

%%%%%%%%%%%%%%%%%%%%%%%%%%%%%%%%%%%%%%%%%%%%%%%%%%

%%%%%%%%%%%%%%%%% BODY OF PAPER %%%%%%%%%%%%%%%%%%

\section{\label{sec:level1} Introduction}

Enhancing our understanding of the formation and evolution of the Universe is strongly dependent upon the observation of molecular matter that exists in the interstellar medium (ISM) and the Solar System. The ISM harbours the materials from which new stars are born, exoplanetary systems are formed, and the building blocks of life are likely created. These processes occur within low-temperature ($<100~K$) high-density regions of space ($10^4~cm^{-3}$) that are mostly opaque to visible radiation \citep{van_dishoeck_2017}. They are, however, observable at longer wavelengths, i.e. throughout the infrared millimetre (mm-wave), microwave and radio wavelength portions of the electromagnetic spectrum. 

The development of related ground-based observational facilities has enabled astronomers to access the spectral ‘fingerprints’ of a huge range of gas-phase chemical species that are of pivotal importance to uncovering the processes in star and planetary formation. During the past decade, for instance, a large-scale facility, the Atacama Large Millimeter and sub-millimeter Array (ALMA). In doing so, however, new questions have been raised associated with understanding the origins and chemical pathways of many large molecules and exotic species identified in star-forming regions (\citep{van_Dishoeck_2014,2011ApJOberg}). This requires increased use of already overcommitted facilities and for which access time is limited through intense user competition, in addition to variable atmospheric conditions. The provision of low-cost and accessible laboratory-based measurement tools that complement observational facilities such as ALMA can provide a very useful alternative for the astrochemist. For instance, by using methods for molecular detection that are akin to those deployed on telescopes, laboratory experiments that emulate the ISM environment can support understanding origins of molecular species and differentiate between gas-phase and solid state chemical pathways. Simulating space environments under fully controlled conditions is thus an attractive experimental capability that offers potential for a step-change approach in performing laboratory astrochemistry.

Our approach towards the above is the development of an experimental technique that is hereafter referred to as Terahertz Desorption Emission Spectroscopy (THz-DES). In essence, the experimental method performs THz (submm-wave) emission spectroscopy of gas-phase molecules that desorb from the solid state in a vacuum environment. Because the molecular density of the desorbed species in the experiment is low, narrow spectral emission features are present, typically a few MHz full-width half-maximum (FWHM), that correspond to those observed within the ISM. We show that by measuring spectral features as the molecules desorb, species can be identified via their excitation states, and kinetic and thermodynamic information such as desorption energies can be extracted \cite{refId0}. Acquiring both spectroscopic and thermodynamic data concurrently provides a novel and valuable laboratory tool to aid in the understanding of astrochemical processes in star-forming regions

Whether deployed on a large single dish telescope (\cite{Davies1992,Little1992}) or interferometric telescope systems \cite{Wootten2009}, heterodyne frequency down conversion systems provide a powerful spectral observation tool. Through its use, more than 200 molecules have been detected and spectroscopically resolved in the interstellar regions within our Galaxy, and more than 60 have been observed in extragalactic molecular clouds. Detections include neutral, radical and ionic or cationic species \cite{MULLER2005215}, and small unsaturated molecules, larger carbon-based molecules dominated by long chain structures with large dipole moments, and fullerenes and polycyclic aromatic hydrocarbons (PAHs) \cite{mcguire20212021}. Predominantly these species are formed by gas-phase chemistry, although it is well known that a number of abundant molecules, such as $H_2$, $H_2O$ and $CH_3OH$ (\cite{1982Tielens,2016ApJwalsh,Ioppolo_2008,refIdFuchs0}) form mostly via solid state pathways involving the dust and ices found in dense interstellar molecular clouds. Vital challenges to astrochemistry are to refine our understanding of the chemical growth mechanisms, and to determine the role that they play in influencing the physical and chemical conditions present as star and planet formation progresses.

One group of species, known as complex organic molecules (COMs) have received significant focus over the past decade from "molecule hunting" astronomers (\cite{Belloche1584,refId0}). COMs are some of the largest molecules detected in space, and since they tend to be saturated organic species, are considered as a bridge between the simpler gas-phase species that are often used to probe or trace astrophysical conditions, and the potential astrobiological processes that may emerge on a newly forming planet or moon. Many of the proposed formation mechanisms for COMs in the ISM derive from processes involving the formation and/or destruction of $CH_3OH$  (\cite{Bergantini_2018,Linnartz2015}). However, it remains unclear whether COMs are formed in the gas-phase or solid state, or a combination of the two; the chemistries may even be molecule dependent. What is clear, though, is that we are only able to detect and identify COMs in space by their gas-phase emission spectra, particularly in the sub-mm and radio regions of the electromagnetic spectrum, where spectral features are species-unique, and given the large dipole moments of many COMs, relatively intense, despite their low abundance (\cite{doiHerbstDishoeck,Balucaniv009}). Moreover, the James Webb Space Telescope (JWST) mission will help understanding the interplay between gas and ice across a variety of different sources in the ISM. A primary objective of the laboratory THz-DES spectrometer described here is, therefore, to detect COMs spectroscopically in the gas-phase once they have desorbed from the solid state and, through a series of experimental variants, concurrently determine the desorption energies of COMs.

To date, laboratory temperature programmed desorption (TPD) experiments (\cite{Fraser2002,Fraser2001,Fuchs2006,Fayolle2011,Collings2004}), using quadrupole mass spectrometry (QMS), have been shown to be of great importance in gaining an understanding of the evolution of ice molecules from dust surface analogues. The TPD technique has historically facilitated the measurement of binding, desorption, and diffusion energies, and the sticking probabilities and desorption orders, of molecules incident upon and released surfaces, and in the context of laboratory astrophysics, proved exceeding productive to astrochemical models and our understanding of simple species in the ISM, e.g. $H_2O$, CO, and $CH_3OH$ (\cite{Fayolle2011,B516770A}). In a few cases (using much higher concentrations of COMs than anticipated in the ISM) the solid state mid-IR spectra and desorption properties of certain COMs have also been reported \cite{refId0Terwisscha} to support direct observation of the ISM.

In the last few years, similar (though not identical) experimental developments to our method have been reported. In 2017, \cite{wheres110} described the design, construction, and operation of two laboratory broadband emission spectrometers for gas-phase characterization of large molecules. The first is based on a Schottky-barrier diode heterodyne receiver spectrometer operating between 80 and 110 GHz, whilst the second uses cryo-cooled SIS technology and operates at higher frequency between 270 and 290 GHz \cite{2018PCCPwehres}. Gas phase pyridine and methyl cyanide features were successfully detected and matching analytic simulations were presented for the observed spectral transitions. In 2018, a new technique combining a Terahertz radiometer (41 - 49)~GHz and a vacuum chamber was developed to observe the generation of cold plasma and UV photochemistry in the gas phase under low pressure as described by \cite{2018Tanarro}. However, in the context of the experiment described here, it is important to note that the spectrometer developments were not coupled to a system where the gas-phase molecules originated from the solid state. More recently, \cite{ycoum} reported on an experiment using a THz source and hot-electron bolometer to detect the gas-phase absorption spectra of simple molecular species such as $H_2O$, $D_2O$ and $CH_3OH$ desorbing into the gas-phase from ices grown in an ultra-high vacuum (UHV) chamber. This experimental set-up (SubLIME), combined with a Fourier-transform Infrared spectrometer, gave the spectroscopic insight on the ultraviolet photolysis and warm up of methanol ice sample, for the first time observed at those frequencies in a laboratory environment \cite{2021yocum}. This showed the potential of the use of submillimetre/far-infrared technique to identify molecules in complex gas mixtures. One challenge of this technique was the sensitivity of absorption spectroscopy to the desorbing molecules. The combination of temperature programmed desorption and microwave spectroscopy techniques was discussed by \cite{doiTheul}. In these experiments, desorption of water, deuterium, methanol, and ammonia was measured using a hot electron bolometer and Si-diode technology combined with Fourier transform spectrometers. While Theule et al. observed the desorption through a waveguide cavity, \cite{ycoum} detected the desorption above a metal substrate as seen in standard TPD studies.

The method presented here differs from other techniques in two ways; first, molecules are detected in emission (not absorption) and second, since an absolute power measurement is made, molecular abundances can be determined. Our method differs notably in that the previous experiments used an illuminating source (100~GHz - 1~THz or 24~GHz) to excite the desorbing molecules. In our technique, we observe the emission spectra against a cold blackbody background that is more representative of the astronomical observation method. We thus measure the absolute intensities from the desorbing molecules. Moreover, the THz-DES bandwidth is much more extensive at 4~GHz per sideband. This allows us to simultaneously observe a range of spectral lines. The THz-DES is therefore able to observe across a full bandwidth with a high acquisition speed of the desorption process without the need for averaging scan and by observing the desorption in real time. Finally, the radiometer used as a detector is compact and operates at room temperature making the system easy to integrate to any UHV chamber, although it has a lower sensitivity when compared with a cooled radiometer system. The aim of this paper is to describe the development and operation of the THz-DES method and to discuss its potential application in the study of COMs desorption from the solid phase. Here we demonstrate the capabilities of the system by using the strong THz emission lines of $N_2O$ as an example of typical operation, showing that both spectra and thermodynamic data can be obtained concurrently during these types of experiments.

\section{THz-DES Instrument Description}
The THz-DES experimental apparatus is designed to perform gas-phase emission spectroscopy of molecules sublimating from interstellar ice analogues through thermal mechanisms. The technique can in principle be also applied to study molecules released from the solid state via non-thermal desorption mechanisms such as photodesorption or ion spattering. By using a broad instantaneous bandwidth semiconductor detector in the form of a Schottky barrier diode mixer, combined with a high-speed digital sampling system, our experiment provides the means of analysing desorption mechanisms whilst producing spectral data resembling direct astrophysical observations of molecules in star forming regions \cite{Mifsud}. A semiconductor device offers the additional advantages of room temperature or cryogenic operation, with the latter offering potential for a future sensitivity enhancement.

\begin{figure*}
\includegraphics[width=\textwidth]{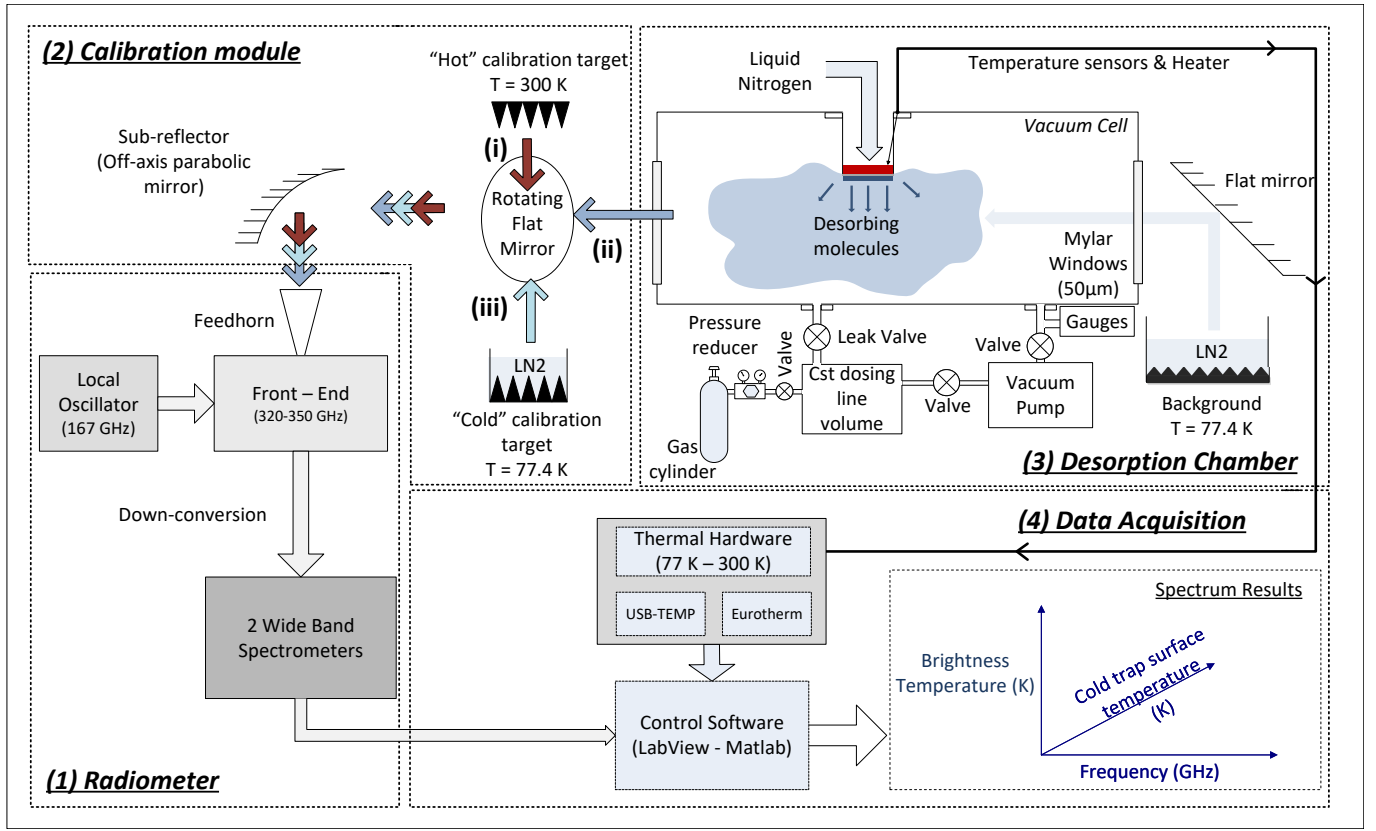}
\caption{\label{fig:RSI}Schematic of the THz-DES experiment with (1) Radiometer; (2) Calibration Module; (3) Desorption Chamber; (4) Data Acquisition. During experiments ices are formed on the cold finger in the vacuum cell by dosing known quantities of gas into the chamber (3). The cold finger is then heated at a controlled rate, measured using platinum resistance thermal sensors. Radiation "arriving" at the front end feedhorn is combined with a local oscillator signal and processed via two fast Fourier transform wideband spectrometers (1). The spectral data (including calibration measurements (2)) and temperature data are combined and processed to generate the final experimental spectra as a function of time (temperature) and frequency with absolute brightness temperature (4).}
\end{figure*}

Our instrument, shown schematically in Figure~\ref{fig:RSI}, comprises a total power heterodyne radiometer (1), a desorption chamber (3) in which the desorption process takes place, a dosing line through which a selected quantity of gas is injected into the cell, and a cold trap that emulates cold surfaces in space such as dust grains and icy planets and moons. The present system provides a measurement capability within the frequency range (320-326.5)~GHz and (340-346.5)~GHz with a spectral resolution and integration time that are user selectable. Spectral images are acquired at a rate that is governed by the integration period selected. The THz-DES system provides absolute intensity measurements of the real line shapes of the molecular features via a calibration module (2).

\subsection{Desorption Chamber}

The molecular chamber cell for this experiment is a high-vacuum system that achieves an ultimate pressure in the range $10^{-6}$~mbar, as measured by Pirani (PR10-K Edwards) and Penning (Penning CP25-K Edwards) gauges. Such a base pressure is sufficiently low to simulate space relevant processes and to allow for the system's demonstration. The cell is composed of a  $1~m$ length of cylindrical steel tubing with an internal bore diameter of $0.13~m$. Selection of the diameter is a compromise between minimising chamber volume to limit molecule-wall interactions post desorption and minimizing the radiometer antenna beam truncation whilst still achieving the maximum beam volume filling and detection sensitivity. The dosing line, cold trap, and a vacuum pumping system are connected via high vacuum ports to the main chamber. A vacuum seal and semi-transparent THz window of $50~\mu m$ thick Mylar (Goodfellow) film is located at both ends of the cell. Each window provides a signal transmission of typically 93\% \cite{Kurtz2005} with negligible absorption within the spectral range of the instrument, but together they introduce a Fabry-Perot interference effect within a signal path that results in a frequency dependent variation of the spectral noise floor. This effect can be to a large extent mitigated through the application of instrument calibration and data post-processing techniques.

\begin{figure}
\includegraphics[scale=0.3]{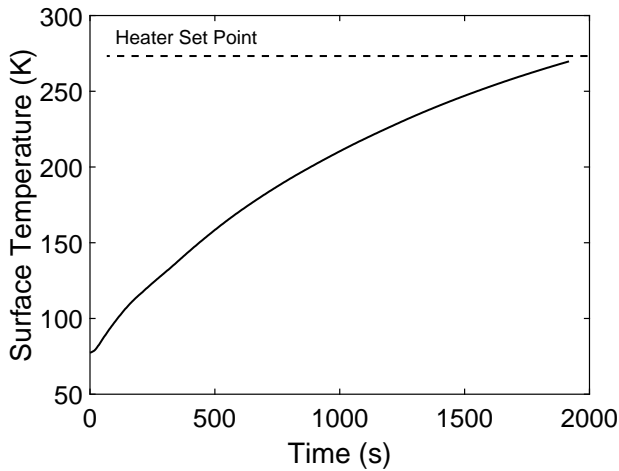}
\caption{\label{fig:ramp} Measured temperature of the cold trap over time during a desorption experiment and with no thermal control.}
\end{figure}

Gas species are introduced into the dosing line via small lecture bottle sources. The line pressure is measured by a Pirani gauge (PR010 Balzers) and thus provides a known and constant volume of gas. Molecular species introduced to the cell are condensed on the cold trap surface, as shown in Figure~\ref{fig:RSI}, which is cooled to liquid nitrogen temperatures. To avoid significant protrusion into the interference with the receiver reception beam, the trap internal diameter and length was restricted to $51~mm$ and $120~mm$, respectively. A co-located heating element and platinum resistance (PRT) thermal sensor are used to respectively vary and monitor the cold trap surface temperature. The heater and thermal sensor are connected to a Eurotherm control unit via a high vacuum feedthrough (LEMO) to achieve a closed loop thermal stabilisation of the trap surface. The cold trap surface temperature is recorded by a computer during a measurement cycle and simultaneously with the spectral data during an experiment. A typical set-point THz-DES temperature curve acquired during a measurement cycle is shown in Figure~\ref{fig:ramp}. When the liquid nitrogen is fully evaporated (point t=0~s in Figure~\ref{fig:ramp}), the temperature of the cold trap begins to increase. By changing the set point, a thermal ramp can be applied thus allowing a degree of control of the rate of temperature increase.

\subsection{Radiometer}

\begin{figure*}
\includegraphics[width=\textwidth]{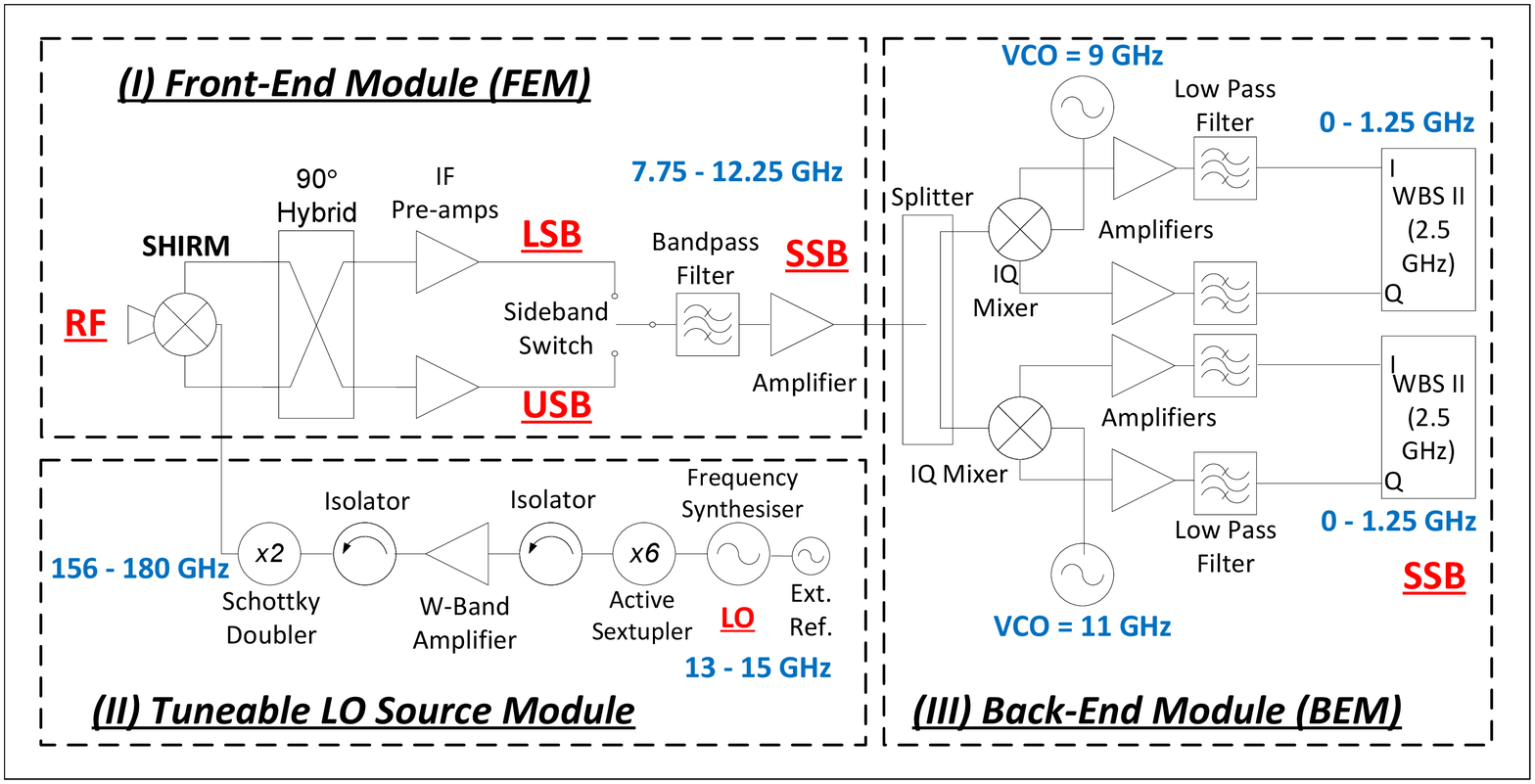}% Here is how to import EPS art
\caption{\label{fig:shirm}Schematic of the heterodyne spectrometer principle. The radio-frequency signal (RF) represents THz emissions. The mixing is taking place within the Front-end module - FEM (I) via the tuneable LO source (II). The signal is then down converted and analysed in the back-end module - BEM (III) with the spectrometer units.}
\end{figure*}

At the core of the THz-DES instrument is a sensitive radiometer system previously developed to measure molecular signatures present within the Earth’s atmosphere \citep{Rea2012_2} and operated at room temperature. The radiometer, which is operated in a total power mode, is divided into two main parts: Front-End Module (FEM), which contains a heterodyne down-conversion receiver system; and Back-End Module (BEM), which performs digital signal processing of the FEM intermediate frequency (IF) down-converted signal. These are outlined in Figure~\ref{fig:shirm}.

The FEM (Figure~\ref{fig:shirm}, (I)) uses a sub-harmonic Schottky barrier diode image rejection mixer in a fundamental mode waveguide structure, which has been described in greater detail elsewhere (\citep{Rea2009,Rea2012}). The sub-harmonic mixer configuration simplifies the provision of local oscillator (LO) power and results in a compact and lower cost instrument front-end.

Mixer sideband separation allows down-conversion of the heterodyne signal bands located either side of the LO into two discrete IF paths that correspond to upper sideband, USB, and lower sideband, LSB \citep{JINSONG} THz frequency inputs. Signal cross-coupling between the sidebands, which is a source of spectral contamination, is typically lower than 30~dB. The IF centre frequency is 10~GHz with a bandwidth of 4.5~GHz per sideband. The baseband frequency of the LO is generated by a microwave frequency synthesizer (XS-14XX, HERLEY-CTI) at (13 to 15)~GHz and combined with an external reference of 10~MHz (OCXO-S109-LF, KVG) to provide a stable output signal. Frequency up-conversion via an harmonic multiplier (x6) and a stage of amplification (APH631, NORTHRUP GRUMMAN) followed by a further stage of harmonic up-conversion (x2) provides a final LO signal of (312 - 360)~GHz, which is injected into the mixer via a waveguide LO input port. In this experiment, the LO was set at 334.5~GHz, which provided a total spectral coverage of (322.25 to 326.75)~GHz and (342.25 to 346.75)~GHz corresponding to the LSB and USB, respectively (Figure~\ref{fig:shirm}, (II)).

After appropriate signal conditioning, i.e. amplification and filtering, the IF output corresponding to each sideband is sampled within the BEM by an 8-bit high-speed analogue-to-digital converter (ADC). A dedicated 2048-point Fast-Fourier transform (FFT) chip then produces a power spectrum of the sampled signal, which, when divided into two 1024-point frames operating in parallel, delivers a real-time sequential flow of spectral data. The combination of signal sampler and processor forms a wideband digital spectrometer (WBS, STAR-Dundee, Ltd), as shown in Figure~\ref{fig:shirm}, (III). Each WBS unit provides an instantaneous spectral bandwidth of 2.5~GHz with a spectral resolution of 1.22~MHz with corresponding spectra stored in a computer for display and further processing. Because of the limited availability of WBS units at the time of development of our instrument, an electronically controlled switch (single-pole double-throw: SPDT) is used to select the desired sideband prior to a filtering stage. This configuration limits the available spectral range of the present system and is a disadvantage that will be rectified as more WBS units become available.

System sensitivity, as defined by the noise equivalent delta temperature ($NE\delta T$) \citep{Hersman1981} depends on various parameters (equation~\ref{nedtformulatheo}) such as the system noise temperature ($T_{sys}$), the spectral resolution (B), the integration time ($\tau$), the calibration procedure ($\tau_c$) and the gain variation ($\Delta G/G$). Use of cooled Schottky (\citep{Zimmerman1991,Siegel998}) or superconducting mixers \citep{Keen1986} would provide enhanced $NE\delta T$ but at the expense of a greater system complexity. This improvement is an intended future instrument enhancement

\begin{equation}\label{nedtformulatheo}
NE \delta T = ( T_{sys} + T_{scene} ) \times \sqrt{ \frac{1}{B \times \tau_{s}} + \frac{1}{B \times \tau_{c}} + \left( \frac{\Delta G}{G} \right)^{2}} .
\end{equation}

Coupling of the molecular emission signal into the mixer is achieved through the combination of an ultra-Gaussian corrugated feed horn antenna, designed by Airbus UK (formerly EADS Astrium), that is directly attached to the mixer, and a subsidiary off-axis parabolic mirror with a focal length of 30.62~mm. The latter directs the signal towards the entrance aperture of the feed horn, which is located at the mirror focus. Viewed from the perspective of the radiometer, the parabolic mirror quasi-collimates the radiometer antenna beam allowing propagation through the cell with minimal truncation and with a Gaussian defined beam waist radius of 6.318~mm as evaluated at a point 1/e from the on-axis amplitude at 320~GHz.

To retrieve the brightness temperature of the observed desorption from acquired data, a linear radiometer transfer function is used \citep{Kerr1997}. The expression of a scene view (line-of-sight) brightness temperature is retrievable depending on the noise contribution in the receiver for the scene view and two reference targets at different temperatures (cold and hot), as explained elsewhere \citep{Kerr1997}. A rotating plane mirror driven by a Brushless DC-Servomotor (FAULHABER) is located between the vacuum cell and parabolic mirror to direct the radiometer input signal path to one of three scene views: (ii) = vacuum cell, (iii) = cold (liquid nitrogen) calibration target and (i) = room temperature calibration target, in Figure~\ref{fig:RSI}. The described instrument calibration procedure, in addition to control of the experimental environmental temperature to within 3K, respectively minimizes the contribution of and variation in background noise. Experimental error arising from background noise is thus mitigated and its influence on the accuracy of the acquired spectral data is considered negligible. The targets provide sources of known radiometric brightness and are thus used to calibrate the radiometric performance of the radiometer. The cold target is formed from a small container lined at its base with a pyramidal microwave absorber (TK RAM, Thomas Keating, Ltd) and filled with liquid nitrogen (assumed physical temperature of 77.4~K at one atmosphere of pressure \citep{1127954}).

The room temperature target, which operates at typically 300~K and for convenience we refer to here as being ‘hot’, consists of an aluminium substrate (65~mm diameter) which is machined in the form of an array of pyramids. A ferrite-loaded epoxy absorber (Emerson and Cuming CR114) is cast onto the aluminium in a manner the replicates the pyramids of the array. The THz surface reflectivity of the target has been previously determined at the required frequency and a power reflection coefficient of less than -30~dB measured across the observational bandwidth. A low reflection coefficient reduces standing waves between the target, receiver and cell windows. It also ensures that the target radiometric brightness temperature is accurately represent by measurement of its physical temperature. The latter is monitored to an accuracy of $\pm0.02~K$ by two thermal sensors that are permanently inserted into the aluminium substrate. In this configuration, the brightness temperature of a molecular species can be compared to the spectral intensity of the same species observed in a molecular cloud, depending on its optical depth and the spectral radiance intensity (source) in accordance to radiative transfer equations and the Rayleigh-Jeans law \citep{198Rybicki}.

\subsection{Radiometer Sensitivity}

The sensitivity of the THZ-DES instrument defines the precision to which the desorption spectral signatures can be measured (the radiometric precision). It is quantified by the minimum change in noise equivalent temperature that can be detected by the radiometer system, commonly referred to as the $NE\delta T$. The parameter is dominated by the inherent electrical noise generated by the radiometer, signal integration time, spectral resolution selection, and radiometer stability and variable spectral baseline artefacts, the latter being dependent upon the environmental and system physical temperature. Performing optimal observations via the THz-DES technique configuration of the instrument parameters (the ‘instrument state’) in order to minimize the $NE\delta T$.

\begin{figure}
\includegraphics[scale=0.32]{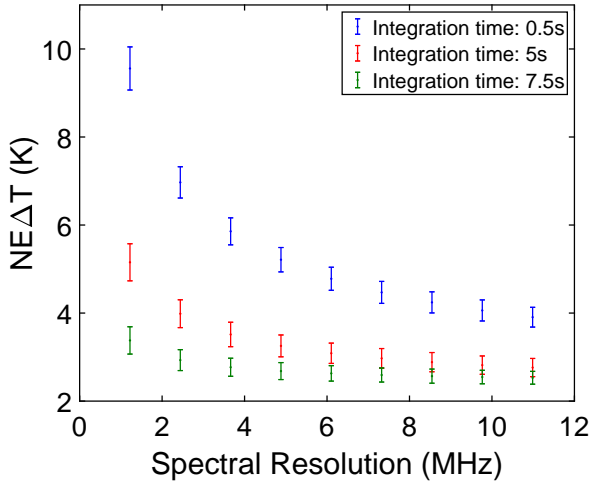}
\caption{\label{fig:NEDT}Measured noise equivalent differential temperature ($NE\delta T$) as a function of the spectral resolution (B) at different integration time ($\tau$). The Blue curve is for 0.5~s integration time whereas the red curve are the results for $\tau$ = 2~s and the green, $\tau$ = 7.5~s.}
\end{figure}

In evaluating the radiometer $NE\delta T$, calculation of the standard deviation of the scene brightness temperature whilst viewing a temperature-stabilised radiometric target in the scene view was performed. From this, the $NE\delta T$ for different integration times and spectral resolutions were measured (Figure~\ref{fig:NEDT}). Consistent with the equation~\ref{nedtformulatheo}, an improvement in radiometric sensitivity is obtained by increasing the spectral resolution (x-axis); an effect that is further enhanced when system integration times are reduced, e.g. 0.5~s (blue curve). However, this sensitivity improvement is at the expense of a degraded spectral resolution capability.

The output noise power of the radiometer system is proportional to the system noise temperature ($T_{sys}$), which is the sum of noise component arising from the antenna, mixers and IF amplifiers. The radiometer operates in Rayleigh-Jeans area of the Black-body spectral region, where the brightness of a Black-body is linearly proportional to the frequency. A sequential introduction of the hot and cold signals into the radiometer input allows the system noise to be evaluated \citep{Dhayes}. When performing the above measurement, both sidebands are evaluated and thus a double sideband (DSB) system noise value, $Tsys_{DSB}$, is determined. In a sideband separating instrument, an adjusted factor is required which governed by the effective sideband gain ratio \citep{Kerr2003}. The typical sideband image rejection factor is respectively $14.8 \pm 1.1~dB$ for the LSB and $20.1 \pm 3.4~dB$ for the USB with a LO at 334.5~GHz and integration time of 1~s. This provides the single sideband (SSB) system noise temperature for each sideband, which, in the case of the radiometric system presented here, is determined to be $4.75\times 10^3 \pm 4.4\times 10^2~K$.

\subsection{System Stability}

The system integration time is dependent upon the nature of the spectral observation undertaken. Variance in the chamber is minimal as the pressure is kept under high vacuum and no changes occurs before desorption. Therefore, it is ultimately limited by the radiometer stability, which is dominated by the system gain drift ($\Delta G/G$). Assessment of the system stability is performed using the Allan Variance measurement method \cite{Schieder2001} which provides a quantitative estimate for the attainable integration time (minima on the Allan curve). This evaluation technique was applied to our THz-DES instrument in which the radiometer viewed a constant temperature radiometric target in the scene view. After an initial stabilization period of 15~mins, the BEM output was recorded at regular time intervals. The Allan Variance was calculated from the raw data using an Overlapping Allan Variance routine \cite{Leupers2013}. The optimum integration can be found when the impact of drift contribution are minimum as well as the lowest white noise level \cite{Schieder2001}. The THz-DES instrument stability becomes compromised with an integration above 80~s \cite{7387275} and any additional integration time does not result in enhanced radiometric precision and beyond this time a recalibration of the system is required. With improved radiometer thermal stability, this limitation can likely be extended. However, the rapid desorption of the ice analogue with increased temperature of the trap limits the integration time to below 80~s.

\section{Experimental Operation}

The instrument LO frequency, digital sampling rate, and integration time can be adjusted via software control to suit user requirements. The system integration time, $\tau$, controls the measurement sensitivity and spectral resolution and the sampling frequency allows the control of the IF bandwidth. Selection of the sideband (LSB or USB) via the electrical switch and tuning of the LO frequency defines the target spectral windows. In addition to influencing the system sensitivity, system integration time defines the rate at which changes in spectral features can be observed. Thus, the instrument set-up needs to be matched to the nature of the experiment. For the current version of the instrument configuration, thermal desorption of a molecular species is a relatively rapid process (typically a duration of 60 s and a rate of 0.1~K/s) and requires a relatively fast acquisition time (typically 0.5~s) in order to capture changing spectral features. As the molecular density in the optical path is low, a high-spectral resolution (< 3~MHz) is required in order to fully resolve the emission line signatures.

\subsection{Ice Deposition}

During the measurement, a defined volume of gas is allowed to enter the cell through the dosing line and with the liquid nitrogen trap cooled to approximately 77.4~K. To minimize species cross-contamination, the dosing line is initially evacuated to a base pressure of approximately $10^{-5}$~mbar via a turbo pump (Leybold vacuum, Turbotronik NT-50), flushed with the selected gas and evacuated again. This procedure is further repeated a minimum of twice before the line is ultimately filled to a pre-determined pressure with the species of interest (25~mbar in 3.2~L volume). If required, multiple species can be added to the cell sequentially by using the same procedure or, alternatively, introduced simultaneously by pre-mixing different gasses within the dosing line. Thus, species can be condensed in pure, mixed and layered form onto the cold trap.

In the experiments presented here, gasses condense onto the cold trap surface to form a solid with an estimated average number of molecules per unit area of order $2\times10^{20}$~molecules/$cm^2$. This estimated value was retrieved from the experimental data, the Ideal Gas Law and the radiative transfer model presented in Section 3.3 (Figure~\ref{fig:BT_P}), by approximating the number of molecules desorbing over the duration of the data acquisition period. This allows the simulation of interstellar and Solar System relevant ice compositions and structures. The THz-DES instrument is particularly suited to investigate hydrogen-deuterium exchanges between, for instance, H$_2$O and D$_2$O \citep{Furuya2016} as well as ortho-/para- transitions \citep{Arasa2015} and symmetry quantum transitions between A- and E- species (e.g. methanol molecules; \cite{DeLucia1989}). Future studies will focus on such topics.

After deposition, the cold trap temperature is increased providing a source of thermal energy that eventually results in a change of state of the molecular material directly from solid to gas phase. The radiometer views the liberated gas through the length of the cell and against a liquid nitrogen (77.4~K) cooled blackbody background target that has been especially designed to present a high emissivity (i.e. close to 1) within the THz frequency range. Because the effective gas temperature is greater than that of the external target, spectral features are recorded in emission against the cold background in a similar manner to observations of an astronomical spectral source.

\subsection{Example Spectrum Signatures}

For our instrument, we have primarily studied N$_2$O. Nitrous oxide has been identified in the ISM \citep{1994Ziurys}, and it exhibits a bright spectral line at 326.556 GHz corresponding to the transition 13(0,0)-12(0,0). Together with other nitrogen oxides, it is also found in planetary environments (\citep{Churchill2000,refId0Ioppolo}) and cometary comae \citep{Saxena2004}. Hence it has been selected as the first study target to evaluate the suitability of our experiment for astrochemical applications.

\begin{figure}
\includegraphics[scale=0.32]{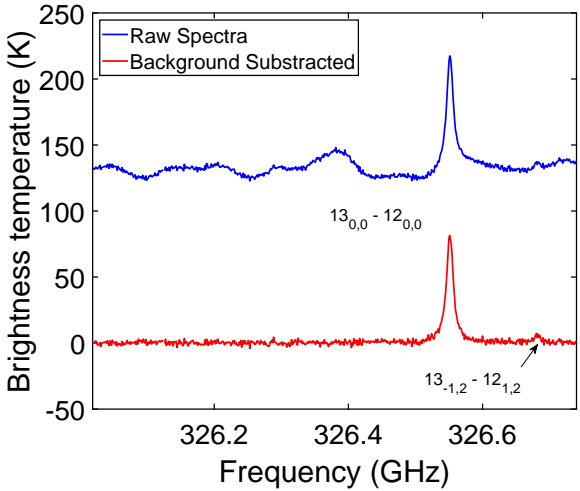}% Here is how to import EPS art
\caption{\label{fig:GENE} Nitrous Oxide spectra at a pressure of $8 \times 10^{-1}$ mbar before (Top) and after (Bottom) background subtraction. Two N$_2$O spectral features are observable at 326.56~GHz and 326.68~GHz after deleted the standing waves coming from the experimental set-up.}
\end{figure}

As shown in Figure~\ref{fig:NEDT}, spectral features can be detected, measured and characterised using the THz-DES system with gas-phase molecules. For experiments with N$_2$O, the local oscillator was set at 334.5~GHz. A background spectral measurement is first performed with the cell at $10^{-6}$~mbar. Multiple reflections between the Mylar windows generate a standing wave that in turn induces baseline artifacts in the power spectrum. The baseline contribution is measured in the absence of the gas and subsequently subtracted from the acquired spectra via data post processing. The background data is then subtracted using a Matlab routine to improve the spectral baseline. A gas-phase spectrum of 0.8~mbar of N$_2$O before (blue curve) and after (red curve) background subtraction is presented in Figure~\ref{fig:GENE}. Two N$_2$O features are easily detectable corresponding to the transition J = 13(0,0)-12(0,0) at 326.55~GHz with a brightness temperature of 80 K above our background temperature and J = 13(-1,2) - 12(1,2) at 326.68~GHz with a peak at 7 K above the background.

\subsection{Desorption Temperature and Molecular Sensitivity}

\begin{figure}
\includegraphics[scale=0.32]{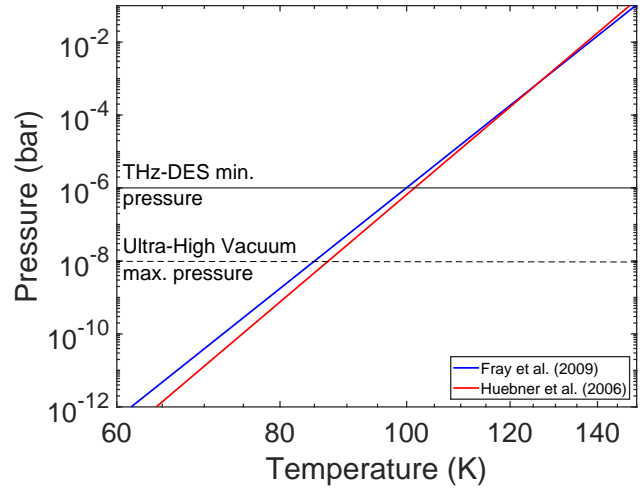}% Here is how to import EPS art
\caption{\label{fig:TempDes} Desorption pressures and temperatures for N$_2$O molecules.}
\end{figure}

Nitrous oxide sublimation has been studied in literature as a function of pressure and temperature (\citep{articleHuebner,FRAY}). Figure~\ref{fig:TempDes} plots past results on the sublimation of N$_2$O and compares them with our instrument's results. Starting at a temperature of 100~K, the THz-DES desorption temperature and pressure ranges are shown above the horizontal black line in Figure~\ref{fig:TempDes}. Figure~\ref{fig:ramp} indicates that the liquid nitrogen cooling system used in the experimental arrangement is appropriate to study desorption of molecules such as N$_2$O.

\begin{figure}
\includegraphics[scale=0.4]{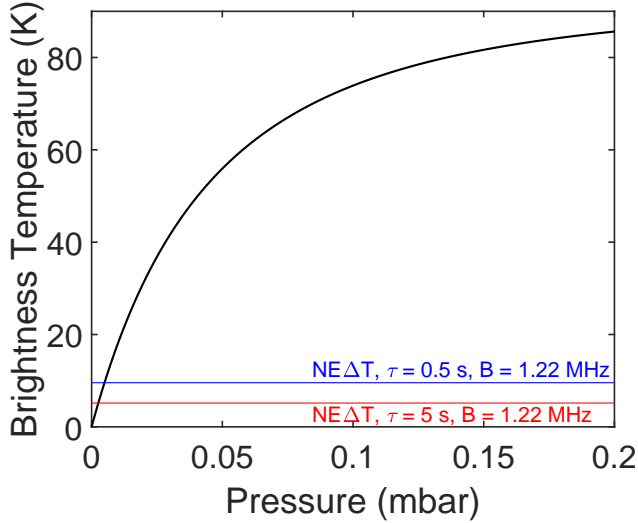}% Here is how to import EPS art
\caption{\label{fig:BT_P}  Brightness temperature of N$_2$O molecules for the transition J = 13(0,0)-12(0,0) at 326.55~GHz as a function of N$_2$O pressure.}
\end{figure}

Figure~\ref{fig:BT_P} shows the brightness temperature of N$_2$O molecules for the transition $13_{0,0}-12_{0,0}$ at 326.55~GHz as a function of N$_2$O pressure. The brightness temperature does not vary linearly as a function of the amount of molecules emitting in the gas phase. The brightness temperature was calculated using line-by-line radiative transfer model with a 1~m gas cell and a blackbody source of 130~K \citep{DUDHIA2017243}. In Figure~\ref{fig:NEDT}, the detection sensitivity threshold is added for integration times of 0.5~s and 5~s, in blue and red, respectively, with a spectral resolution of 1.22~MHz. The THz-DES results presented in this manuscript were acquired with the settings corresponding to the blue threshold, which in the case of N$_2$O detection is $4 \times 10^{-3}$~mbar. Due to the relatively high pressure in the chamber, the desorbed molecules are considered to be in thermodynamic equilibrium with the chamber wall when they are detected by the radiometer.

Other technologies such as QMS, THz bolometer \cite{ycoum}, and Si-diode \cite{doiTheul} instruments, offer a higher sensitivity when compared with the current iteration of the THz-DES. However, our THz-DES approach provides observational advantages through direct measurement of the emission lines of molecules as they desorb, and in real time and over a large instantaneous bandwidth. This allows the characterisation of each desorption process in the chamber, independently. Additionally, it is able to detect a wider range of species within its frequency range such as radicals, or complex molecules and which are hard to detect via other techniques. Figures~\ref{fig:TempDes} and \ref{fig:BT_P} show that the current configuration for our THz-DES system is fully operational under high-vacuum conditions. The future introduction of UHV conditions in the gas cell will require the use of a lower surface temperature and enhanced radiometric sensitivity. The first can be achieved with a different cryogenic system onto which deposit ices. Due to the lower pressure conditions in an UHV chamber, the desorption temperature for $N_2O$ is between 60 and 80~K, requiring the use of, for instance, a helium close-cycle cryostat to deposit it onto a cold substrate. Moreover, to reach UHV conditions, a better pumping system is required because the effective pumping speed of the THz-DES gas cell is measured around 20 L/s, while standard UHV systems are greater 100~L/s. Finally, enhanced radiometric sensitivity could be achieved by cryo-cooling the receiver mixer low-noise IF amplifier which, although adding instrument complexity and cost.

\section{DESORPTION RESULTS}

\subsection{THz-DES Emission Spectra}

\begin{figure}
\includegraphics[scale=0.3]{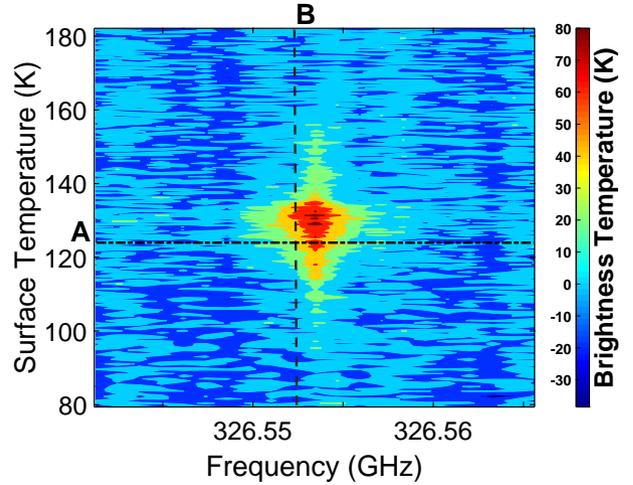}% Here is how to import EPS art
\caption{\label{fig:3D} N$_2$O THz-DES Result for one exposure time deposition. The contour figure of the brightness temperature of desorbed N$_2$O molecules is plotted as a function of the frequency (x-axis) and the trap physical temperature (y-axis). The A-line is a cut of the brightness temperature as a function of the frequency at $T_{surface}$ = 124~K corresponding to the red curve in Figure~\ref{fig:CutF}. The B-line is a cut of the brightness temperature as a function of the surface temperature at 326.551~GHz corresponding to the green curve in Figure~\ref{fig:CutT}.}
\end{figure}

In a first experiment, N$_2$O was allowed to condense onto the cold trap at 77.4~K. By heating the trap, the condensed material was then thermally desorbed. Figure~\ref{fig:3D} shows THz-DES spectra as a function of the surface cold finger temperature (K)($T_{surface}$) and brightness temperatures (K). Changes in the spectra profiles observed in Figure~\ref{fig:3D} are consistent with variations in molecular density. For instance, as desorption increases, the molecular density within the cell environment rises producing a corresponding increase in brightness temperature and linewidth at the spectral line frequency. The onset of N$_2$O desorption appeared to occur at $T_{surface}$ of approximatively 100~K, as measured by Fray et al.\citep{FRAY}. At $T_{surface}$ = 129~K, all N$_2$O molecules have been desorbed and the maximum intensity is observed. The molecular density within the cell then gradually decreases towards zero as molecules exit the cell via the vacuum pump resulting in a corresponding reduction in spectral linewidth and brightness temperature (from $T_{surface}$ = 159~K).

\begin{figure}
\includegraphics[scale=0.32]{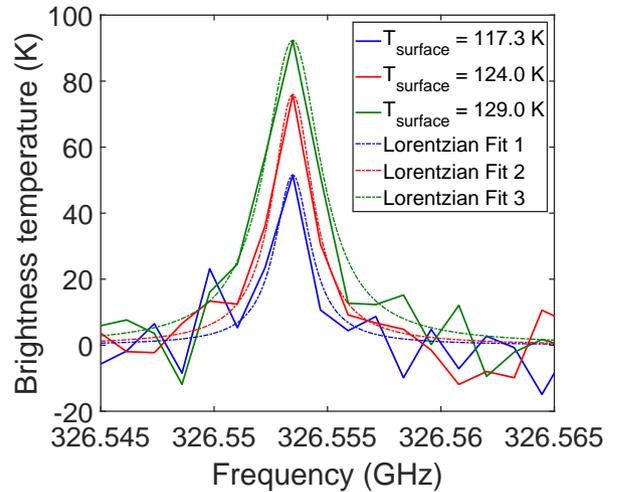}
\caption{\label{fig:CutF}Brightness temperature for spectral feature of N$_2$O molecules as a function of the frequency at different trap physical temperature, respectively 94.8~K (blue curve), 115.3~K (red curve), 129.0~K (green curve) and 141.5~K (black curve).}
\end{figure}

The brightness temperature for different $T_{surface}$, as a function of WBS frequency channels (A-line in Figure~\ref{fig:3D}), is shown in Figure~\ref{fig:CutF}. Brightness temperatures for $T_{surface}$ of 94.8~K (blue curve), 115.3~K (red curve), 129.0~K (green curve and previous A-dashed line) and 141.5~K (black curve) are shown from 326.545~GHz to 326.565~GHz. At $T_{surface}$ = 94.8~K, no N$_2$O is desorbed showing the noise floor of the experiment. Desorption commences as the temperature further rises above 100 K. The brightness temperature increases at the frequency of the N$_2$O feature (326.553 GHz) as the number of molecules emitted is higher. The value of the brightness temperature rises from 56.74~K to 92.36~K when $T_{surface}$ increases from 115.3~K to 129~K. When the gaseous molecules are removed from the vacuum cell, the brightness temperature decreases (40.98 K at $T_{surface}$ = 141.5~K) and eventually returns to the THz-DES noise floor.

The line broadening effect is observable in the frequency domain (as seen in Figure~\ref{fig:CutF}). Adjacent spectrometer channels to the centre line frequency, 326.553~GHz for N$_2$O, are tracers of this broadening effect. At first, natural broadening occurs as the number of molecules desorbing in gas phase is low. Then the spectral line FWHM increases due to pressure broadening \citep{1987Chamberlain}, which results in a time and temperature dependent spread function across different spectrometer channel frequencies. The resulting line-shape function is a Lorentzian (dashed-lines in Figure~\ref{fig:CutF}; \cite{Tanssi}). The spectral line intensifies and becomes wider over the frequency (red vs. green curves). The line intensity increases until the molecular density in the beam path reaches a value where self-broadening is important, leading to a higher FWHM, from 1.9~MHz to 3.0~MHz at $T_{surface}$ of 117.3~K and 129.0~K.

\begin{figure}
\includegraphics[scale=0.32]{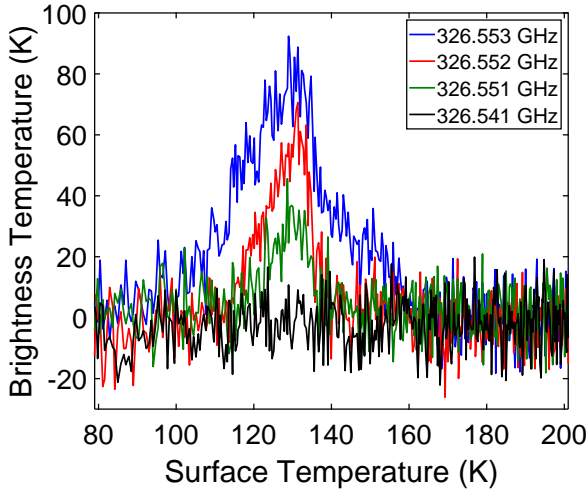}
\caption{\label{fig:CutT} Brightness temperature for spectral feature of N$_2$O molecules for its peak frequency (blue curve) and the “wings”, which are the nearest spectrometer channels (red, green and black curves).}
\end{figure}

Another way of analysing the measured brightness temperature of the desorbed N$_2$O is to plot it against $T_{surface}$ for specific frequency channels (B-line in Figure~\ref{fig:3D}) as shown in Figure~\ref{fig:CutT}. The four frequency channels correspond to the central frequency line channel at 326.5535~GHz (blue curve) and at the following spectral intervals from the central frequency: 1 channel (red curve), 2 channels (green curve) and 5 channels (black curve). The red and green traces represent tracers of pressure broadening, as stated above, from which it is observed that the intensity increases until the molecular density in the beam path reaches a value where self-broadening becomes apparent. By analysing the central frequency, N$_2$O desorption starts around 100~K, i.e. when the intensity from the blue curve becomes higher than the noise floor. From $T_{surface}$ = 95 - 135~K, the brightness temperature increases to reach a maximum value at approximately 80~K. At this point, more N$_2$O ice is sublimated into the desorption chamber. The newly gaseous molecules emit with more intensity as their amount increases. The brightness temperature provides a direct measurement of this phenomena. From $T_{surface}$ = 130 - 160~K, the pumping rate becomes higher than the desorption rate. The number of molecules in the chamber drops, as observed from the brightness temperature. After 160~K, no radiation is detected. All icy molecules have desorbed and are pumped and hence removed from the cell.

\subsection{Water Contamination Impact}

\begin{figure}
\includegraphics[scale=0.32]{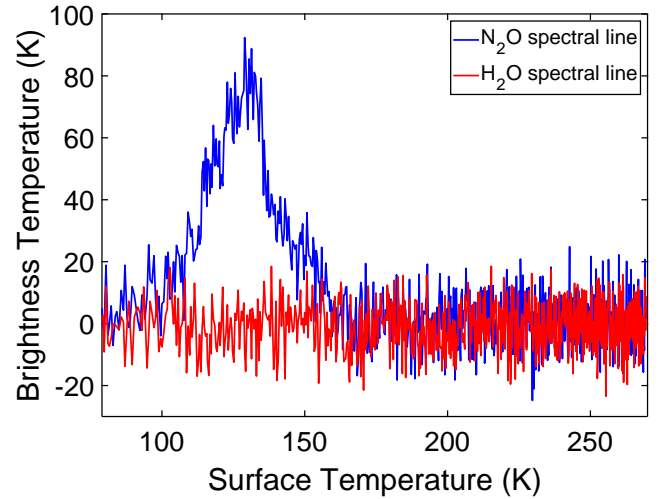}% Here is how to import EPS art
\caption{\label{fig:n20_h20_des}  Brightness temperature of N$_2$O molecules at 326.56~GHz (blue curve) and brightness temperature of possible H$_2$O molecules at 325.15~GHz (red curve).}
\end{figure}

Under our current experimental conditions, water molecules can co-deposit with the nitrous oxide during gas dosing and throughout the ice experiment for a cold trap temperature below the desorption of water ice. The THz-DES is able to monitor the presence of any molecule which has a rotational spectral feature within the THz-DES frequency range. Fortunately, this is the case for water which presents a para-$H_2O$ $5_{1,5} - 4_{2,2}$ at 325.153 GHz \cite{PICKETT1998883}. Water is expected to either co-desorb with $N_2O$ via a volcano mechanism \cite{PhysRevLettSmith} or desorb from around 200~K at a pressure of $10~^{-6}$~mbar \cite{FEISTEL200736}. Figure~\ref{fig:n20_h20_des} shows the desorption of the nitrous oxide (blue curve) and the brightness temperature corresponding to the frequency of the water line at 325.153 GHz (red curve). No water desorption is detectable for any surface temperature, either while the nitrous oxide is desorbing or when the water desorption temperature is reached (200~K). This shows that water contamination is negligible due to the relatively short deposition and desorption experiment that takes place within a period of 30 - 45 minutes.

\subsection{Ice Thickness Dependence}

\begin{figure}
\includegraphics[scale=0.32]{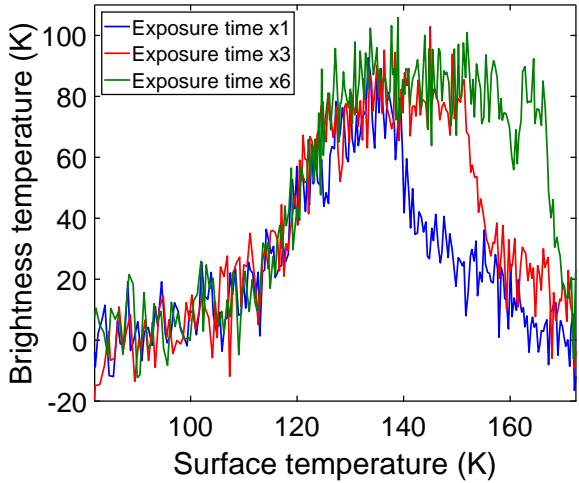}% Here is how to import EPS art
\caption{\label{fig:Thick} Brightness temperature versus surface temperature as a function of the ice thickness for N$_2$O. The blue curve is related to the brightness temperature for an ice made via 1 exposure time deposition whereas in the case of the red curves, 3 times more N$_2$O has been deposited and for the green curve 6 times more.}
\end{figure}

According to TPD experiments reported in literature \citep{Fraser2002}, the thickness of a desorbing ice (in multi-layer regime) should not impact the desorption rate because only the order of reaction would impact the desorption slope \citep{doiThrower}.
The growth of an ice layer of an estimated solid phase molecular density of $2 \times 10^{20}$ molecules/$cm^3$ is referred to as one-deposition. Ices with three times this deposited amount and 6 times larger than the first one have been studied independently with THz-DES. Figure~\ref{fig:Thick} compares the brightness temperature intensity measured for the central feature frequency for one 'deposition' (blue curve), three 'deposition' (red curve) and 6 'deposition'(green curve). The brightness temperature maximum is reached when an optical density saturation effect is reached. The three N$_2$O thick ice layers start to desorb at the same temperature (around 90~K) and follow the same desorption trend, from $T_{surface}$ = 90 - 125~K. However, the desorption lasts longer for the thicker ice as expected due to a larger amount of condensed matter desorbing (red and green curves), at $T_{surface}$ of 140~K (blue curve), 155~K (red curve) and 170~K for 6 exposure times deposition (green curve). As the pumping rates are similar in all experiments, the gradient of the respective brightness temperatures is also similar. Therefore, the thickness of the N$_2$O ice does not impact the thermal desorption rate, which is in good agreement with previous TPD experiments of similar thermal processes.

\subsection{Thermal Desorption Activation Energy}

In a thermal desorption experiment, the activation energy corresponds to the energy required for the pure ice to change phase from solid to gas. The thermal desorption activation energy ($E_{des}$) can be measured with standard TPD techniques and is usually retrieved from the Polanyi-Wigner equation and Arrhenius equation, described in detail elsewhere \citep{DEJONG1990355}. The rate of desorption over temperature($R_{des}$), shown in equation~\ref{equR} \citep{KING1975384}, depends on a pre-exponential factor ($\nu(\theta)$), the surface coverage ($\theta$), the kinetic order of reaction (m), the heating rate ($\beta$), the gas constant (R), the desorption temperature (T) and desorption activation energy ($E_{des}$)

\begin{equation}\label{equR}
   R_{des}(T) = \frac{\nu(\theta) \times \theta^m}{\beta} \times exp \left(-\frac{E_{des}(\theta)}{R \times T}\right).
\end{equation}

Only the desorbing region ($T_{surface}$ = 90 - 125~K) in Figure~\ref{fig:Thick} are used to quantitatively determine the N$_2$O thermal desorption activation energies for the three ice thicknesses. To calculate $E_{des}$, each brightness temperatures are fitted using the Reference Forward model (\cite{DUDHIA2017243}; shown in Figure~\ref{fig:BT_P}) in order to retrieve the amount of molecules desorbing at their corresponding surface temperature. As stated previously, the desorption's slopes in Figure~\ref{fig:Thick} are similar for various surface coverage. This corresponds to a $0^{th}$ reaction kinetic order and would made the term $\theta^m$ equal to 1 in equation~\ref{equR}. The THz-DES desorption activation energies calculated for one, three and six deposition exposure times are $13.4 \pm 3.2$~kJ/mol, $10.9 \pm 0.3$~kJ/mol and $13.9 \pm 2.5$~kJ/mol, respectively. In the literature, $E_{des}$ for N$_2$O ices is between 18 and 29~kJ/mol with standard TPD experiments \citep{TF9302600196,doiBryson,CORNISH1990209,doiKiss,SAWABE199245,10Brown,doiLian,doiToker}. The reference enthalpy energy for sublimation of N$_2$O is given between $23.6 \pm 0.4$~kJ/mol \citep{doiBryson}. This value correspond to the sublimation of a N$_2$O of an ice of more than $ 10^{19}$ molecules with a pressure up to $ 10^{-4}$~mbar.

The thermal desorption activation energies measured with the THz-DES experiment are therefore within the same order of magnitude of those presented in the literature. Differences could be explained by limitations within the THz-DES experimental set-up, such as low pressure (only high-vacuum), possible re-adsorption of desorbed molecules, substrate made of stainless steel with Kapton heaters, non-linear heating rate, etc. However, it is also possible that the measured desorption activation energies is accurate and would simply differ from literature values because of the existence of spectroscopic preferential desorption mechanism, called rotational cooling. For instance, this was measured by \cite{doiLian}, where the N$_2$O desorption energy was about 22.19~kJ/mol at 2236~$cm^{-1}$ and 28.95~kJ/mol at 2270~$cm^{-1}$. To validate this hypothesis, it would be necessary to characterise the THz-DES signal of different spectral transitions to probe the presence of rotational cooling, which is beyond the scope of this work.

\section{Conclusions}

We have developed an experimental method that emulates thermal desorption mechanisms that occur in star forming regions of the ISM and that can therefore be used to support astrochemical investigations. Our initial instrument configuration uses a heterodyne receiver operating between 320-350~GHz in a sideband separating detection mode with interfacing optics and a calibration system that together form a sensitive radiometer operating at room temperature. The system is combined with a vacuum desorption cell, dosing line, and cold trap to freeze and then thermally desorb species from the solid to the gas phase that are then spectroscopically observed against a cold background reference. Spectral emission brightness temperatures of species are recorded over time with high spectral resolution.

The THz-DES system exhibits the advantage of measuring the absolute emission intensities of desorbing molecules by calibrating the measured spectra. The acquisition is performed in real time, that is no averaging scan, with a 4~GHz, single-sideband, instantaneous bandwidth. The spectroscopic information provided by the THz-DES experiment allows the identification of molecules desorbing from an ice and emitting against a 77.4~K blackbody in the spectral range covered by the instrument. This specific instrumental feature is potentially advantageous for the study of radicals and complex organic molecules, since their unambiguous identification in standard TPD measurements with a QMS are not straight forward.

Our work has demonstrated the system performance and its potential future impact in astrochemical studies of thermal and non-thermal desorption mechanisms. We have shown, for instance, that desorption of different thicknesses of $N_2O$ ice does not introduce changes in the desorbing rate of $N_2O$. The thermal desorption activation energy for $N_2O$ was calculated from the THz-DES results between 10 and 18~kJ/mol, which correlates with previous results based on standard quadrupole mass spectrometry techniques as reported in the literature (18 to 29~kJ/mol). Thermal desorption would appear to be a rotational energy selective mechanism as the activation energies vary depending on the spectral lines observed. By observing different spectral lines with THz-DES for one molecular kind, it will be possible to identify any rotational cooling during desorption \cite{TULLY1994667}.

Future instrumental developments will include the introduction of a wider spectral bandwidth through the addition of more radiometer channels to increase the spectral observation range. An improved sensitivity will be achieved by cooling the FEM mixer and the addition of more WBS units to allow simultaneous observation. When combined with an ultra-high vacuum gas cell, enhanced simulation of the interstellar and Solar System environments will be possible. This will support astronomers in the investigation of a series of thermal and non-thermal desorption processes that include photo and chemical desorption, and cosmic ray induced sputtering of the molecular ice.

\section*{Acknowledgements}

The authors would like to thank the support of Star Dundee, Ltd for its advice and help regarding the WBS spectrometer software. SI acknowledges the Royal Society for financial support.

%%%%%%%%%%%%%%%%%%%%%%%%%%%%%%%%%%%%%%%%%%%%%%%%%%
\section*{Data Availability}

The data that support the findings of this study are available from the corresponding author upon reasonable request.



\bibliographystyle{mnras}
\bibliography{aipsamp} % if your bibtex file is called example.bib

%\begin{thebibliography}{99}
%\end{thebibliography}

%%%%%%%%%%%%%%%%%%%%%%%%%%%%%%%%%%%%%%%%%%%%%%%%%%

%%%%%%%%%%%%%%%%% APPENDICES %%%%%%%%%%%%%%%%%%%%%

%\appendix

%%%%%%%%%%%%%%%%%%%%%%%%%%%%%%%%%%%%%%%%%%%%%%%%%%

% Don't change these lines
\bsp	% typesetting comment
\label{lastpage}
\end{document}